# A HYBRID METHOD OF ACCURATE CLASSIFICATION FOR BLAZARS OF UNCERTAIN TYPE IN FERMI LAT CATALOGS

Yijun Xu,[1,2,3] Weirong Huang,[1,2,3] and Hui Deng[1,2,3]

—

Ying Mei[1,2,3] and Feng Wang[1,2,3]

[1]*Center For Astrophysics, Guangzhou University*
*Guangzhou,Guangdong, China, 510006*
[2]*Astronomy Science and Technology Research Laboratory of Department of Education of Guangdong Province*
*Guangzhou,Guangdong, China, 510006*
[3]*Key Laboratory for Astronomical Observation and Technology of Guangzhou*
*Guangzhou,Guangdong, China, 510006*

## ABSTRACT

Significant progress in the classification of Fermi unassociated sources , has led to an increasing number of blazars are being found. The optical spectrum is effectively used to classify the blazars into two groups such as BL Lacs and flat spectrum radio quasars (FSRQs). However, the accurate classification of the blazars without optical spectrum information, i.e., blazars of uncertain type (BCUs), remains a significant challenge. In this paper, we present a principal component analysis (PCA) and machine learning hybrid blazars classification method. The method, based on the data from Fermi LAT 3FGL Catalog, first used the PCA to extract the primary features of the BCUs and then used a machine learning algorithm to further classify the BCUs. Experimental results indicate that the that the use of PCA algorithms significantly improved the classification. More importantly, comparison with the Fermi LAT 4FGL Catalog, which contains the spectral classification of those BCUs in the Fermi-LAT 3FGL Catalog, reveals that the proposed classification method in the study exhibits higher accuracy than currently established methods; specifically, 151 out of 171 BL Lacs and 19 out of 24 FSRQs are correctly classified.

Corresponding author: Feng Wang fengwang@gzhu.edu.cn

1. INTRODUCTION

Blazars, one of the subclasses of radio loud quasars, are considered a type of energy source that emit their relativistic jet directly at the observer. Blazars have a series of interesting observational features: high variability, high polarisation, and high redshift. However, the radiation mechanism inside these objects is still an open question and the origin of the gamma sources remains a mystery. The classification of quasars has always been a prerequisite for astronomers to study these objects. Blazars can be divided in to two subclasses: BL Lacs and flat spectrum radio quasars (FSRQs). Observationally, BL Lacs have a lower luminosity than FSRQs. The most common and preferred method of classifying a blazar is verification of its spectrum, which comprises nonthermal radiation in the domain ranging from radio to gamma wavebands. However, observational features in the spectrum of BL Lacs differ from those of FSRQs. In the optical spectrum, specifically, FSRQs have a flatter spectrum with stronger (wider) lines, whose equivalent width (EW) $\geq 5°A$, whereas BL Lacs have a steeper spectrum and with weak or even no lines, which may indicate that two different physical processes are occurring within them.

An alternative, is multiwavelength analysis which has been extensively used recently. There are two humps in the multiwavelength spectral energy distribution in the $log\nu - \nu log\nu$ space. One hump is in the domain of synchronic radiation in a low energy band, and the other is in the domain of inverse-Compton scattering in a high energy band. The location of the lower energy hump can serve as the second standard for classifying blazars: BL Lacs tend to have a higher peak frequency, whereas FSRQs have a lower peak frequency, $F_{pfsrq} \leq 31$, $F_{pBLLac} \geq 5783$ (Kang et al. 2019)
.

However, there is an overlap in the distribution of the peak frequencies of different sources, making it difficult to classify the sources. Furthermore, there are certain limitations, i.e. the spectra of certain sources or their optical counterparts cannot be obtained. In addition, these two methods both require completely visual identification processes. This leads to biases and requires considerable manpower.

All the differences in observation between BL Lacs and FSRQs suggest that there may be two different inner physical origins and radio mechanisms in these two objects which should be studied separately. To study these two objects, more high quality observational data and new technical methods are required.

After the launch of the Fermi satellite, many gamma ray sources were detected, and a large number of them were found to be blazars. Not only did the number of detected blazars increase, the improved sensitivity also provides a good chance for identifying and locating the sources. However, numerous sources were labelled as blazars of uncertain type (BCUs) in the third release of Fermi satellite data (3FGL) (Acero et al. 2015). In addition, the ratio of BCUs increased during the process of survey, from 13.8% in 1FGL to 33.4% in 3FGL (Chiaro et al. 2016), making extensive optical observation for those BCUs difficult and taxing. Thus, researches on the BCUs with Fermi data have gain its popularity (i.e. Salvetti et al. (2017) , Hassan et al. (2012) , Lefaucheur & Pita (2017)). Furthermore, classifying BCUs can provide potential data for statistical research on blazars.

As there are limitations in the observation process, and blazars classification methods that do not fully depend on obtaining an optical spectrum have been developed. One such method provides classification through multiple features, representing multiple wavelength examination (Einecke (2016), Fermi data are compared with infrared and x-ray data).

Another method gives a classification result (namely a possibility) directly based on observational data through machine learning. Currently, machine learning and data mining are powerful and effective tools ((Ball & Brunner 2010)). Numerous studies have utilised these methods in recent years, e.g. the classification of sources in the second data release by Gaia (Bai et al. 2018) and the study of dark matter by Mirabal et al. (2012). Kang et al. (2019) and Parkinson et al. (2016) attempted to classify BCUs in 3FGL via machine learning using the R language.

However, in nearly all previous work, result were evaluated only by comparing them with their test set, rather than their real classification labels. In the fourth release of Fermi satellite data (4FGL) (Abdollahi 2019), some BCUs in the 3FGL Catelog have been spectrally determined. This provides a very good opportunity for diagnosing the usability and accuracy of the BCU classification methods, which serve as reference for future study in this field. In this work, we focus on the study of a robust classification method. We compare the classification results with the labels given in 4FGL and evaluate the accuracy, which gives a more reliable assessment of our methods.

The rest of the paper is organized as follows. In section 2, the data source has been briefly introduced. The data pre-processing and classification method are detailed in section 3 in depth. The results are shown and discussed in section 4. Finally, section 5 concludes the paper and present certain aspects of our future work.

## 2. DATA SOURCES AND PREPARATION

### 2.1. *Data sources and preparation*

3FGL contains the results of the gamma waveband survey over a four-year period. It consists of three main gamma source labels: AGN(Active Galactic Nuclei), pulsars, and other gamma sources that cannot be classified. Among AGN, 1144 blazars are labelled as either BL Lacs or FSRQs, and 573 are labelled as BCUs. Uncertainty and other calibration information are not considered in this work. Therefore, each blazar is described by 14 features including its power law index and other radiative information.

### 2.2. *Data screening*

The raw data for all 1144 labelled blazars, comprising 660 BL Lacs and 484 FSRQs, are considered for machine learning. Each source has 14 features, which are shown in Table 1. Each column represents a different feature for a single source; different sources have different values for the same feature (each row). The flagged sources in 3FGL are eliminated to improve accuracy, resulting in 604 BL Lacs and 414 FSRQs remaining.

Table 1. Example of 3FGL data with 14 features

| Source Name | 3FGL J0001.2-0748 | 3FGL J0001.4+2120 | 3FGL J0004.7-4740 | ... | 3FGL J2359.5-2052 |
|---|---|---|---|---|---|
| ROI num | 323 | 326 | 435 | ... | 751 |
| Signif Avg | 11.253237 | 11.350013 | 14.519591 | ... | 6.9364586 |
| Pivot Energy | 1263.8359 | 310.68167 | 722.59973 | ... | 2371.264 |
| Flux Density | 4.85E-13 | 2.52E-11 | 2.13E-12 |  | 6.06E-14 |
| Flux1000 | 6.95E-10 | 2.94E-10 | 6.93E-10 | .. | 3.37E-10 |
| Energy Flux100 | 7.82E-12 | 8.07E-12 | 9.19E-12 |  | 3.75E-12 |
| Signif Curve | 1.9002377 | 4.4135795 | 1.835987 |  | 0.13524954 |
| Spectral Index | 2.1493483 | 2.3051133 | 2.4035163 |  | 2.0176828 |
| PowerLaw Index | 2.1493483 | 2.7774532 | 2.4035163 |  | 2.0176828 |
| Flux100 300 | 4.09E-09 | 1.52E-08 | 1.11E-08 | .. | 4.57E-09 |
| Flux300 1000 | 2.08E-09 | 3.64E-09 | 3.38E-09 |  | 5.63E-10 |
| Flux1000 3000 | 5.22E-10 | 3.57E-10 | 6.01E-10 |  | 1.99E-10 |
| Flux3000 10000 | 2.11E-10 | 8.41E-15 | 1.34E-10 | ... | 5.92E-11 |
| Flux10000 100000 | 1.01E-11 | 1.04E-14 | 4.87E-16 | ... | 5.46E-11 |
| Variability Index | 49.738213 | 130.33633 | 112.932655 | ... | 36.292473 |
| CLASS1 | bll | fsrq | fsrq | ... | bll |

Note—$2_{nd}$ to $15_{th}$ rows show the features used as learning material; $16_{th}$ row shows the given classification result used as a label. Those sources labelled as BLL and bll are both considered as BL Lac, and sources labelled as FSRQ and fsrq are both considered as FSRQ in this work.

## 3. DATA PRE-PROCESSING AND METHODS

The aim of this work is to classify BCUs into BL Lacs or FSRQs through machine learning. Supervised machine learning (SML) is the most common method of building a classifier in this field. SML trains a classifier by training data containing features and a label and then classifies unknown data using only their features. In our work, 1018 labelled blazars were used as the learning material, which is a small sample in SML. The machine learning algorithms employed in this work are described below.

### 3.1. *Algorithms and data preparation*

#### 3.1.1. *Algorithms*

Based on the previous literature on machine-learning models for astronomy, six classical algorithms were selected for use in the study: RandomForest (RF)(Breiman 2001), Multilayer perceptron (MLP) (Rumelhart et al. 1985), Support vector machine (SVM) (Vapnik 2013), Decision tree (DT) (Utgoff 1989), K-nearest neighbours (KNN) (Arthur & Vassilvitskii 2006) and Logistic regression (LR).We directly used the specified functions in the *scikit-learn* (Pedregosa et al. 2011) package to build the classifiers.

All the SML models are the basic type, and the related parameters for each model are listed in Table 8. Some traditional classifiers such as decision tree are applied in this study, the same or improved classifiers are applied to the same data(3FGL). For another aspect, a more comprehensive comparison between the hybrid method and the previous one can be delivered on the traditional classifiers and the later ones(i.e. Random Forest).

#### 3.1.2. *Data preparation*

SML algorithms require a division of the entire learning sample into a training set and a testing set. The data for each blazar consist of features, which are used for learning, and a label, which represents classification results, i.e. BL Lac or FSRQ. Of all labelled blazars used in this work, 80% are chosen to constitute the training set and 20% constitute the testing set. During the training process, the training set is utilised to train the model (i.e. build a classifier). Once the training is completed, the testing set is used to test the model's performance, which is reported as accuracy. Once the model is trained and passes the test, it is applied to predict the class in which BCUs belong.

### 3.2. *Preliminary analysis*

We trained several models using all the aforementioned observational features and used them as a reference. The accuracy of each model is provided in Table 2. We obtained a higher accuracy in SVM(linear), in LR, and in MLP(LBFGS), respectively, indicating that these machine learning algorithms can classify blazars accurately.

However, when the parameters for the models were optimised, it was observed that the models required considerable computing time, as shown in Table 2. In particular, MLP with the L-BFGS kernel required 379 s. We concluded that this issue was caused by the inclusion of a large number of features in the computing process. This suggests that certain features that have only a weak connection with the result may be included in the modelling process, resulting in invalid calculation and eventually increasing the computing time.

Table 2. Accuracy and elapsed time of each model before/after deploying a PCA algorithm

| Model | Accuracy before PCA | Elapsed time before PCA(s) | Accuracy after PCA | Elapsed time after PCA(s) |
|---|---|---|---|---|
| RF | 91.28% | 185.27029 | 91.28% | 182.22628 |
| SVM (linear) | 90.39% | 161.56140 | 89.45% | 88.39529 |
| SVM (rbf) | 61.14% | 0.46908 | 61.14% | 0.08131 |
| SVM (sigmoid) | 61.14% | 0.29905 | 72.93% | 0.02878 |
| MLP (lbfgs) | 91.70% | 379.84447 | 92.20% | 153.45695 |
| MLP (sdg) | 64.19% | 22.03468 | 86.46% | 24.67881 |
| LR  | 86.90% | 85.50223 | 90.83% | 6.38449  |
| DT  | 82.53% | 7.38238  | 86.70% | 5.36331  |
| KNN | 87.34% | 17.45031 | 87.34% | 14.72359 |

### 3.3. Principle Component Analysis and Machine Learning

To reduce computing time, we require a method that can ensure accuracy while eliminating the features that have a weak connection with the classification result.

#### 3.3.1. Evaluation of the contribution of each feature

Typically, features with a large contribution to the label (classification result) are used in machine learning. A similar method was used in another work: Kang et al. (2019) combined the result obtained from the Kolmogorov–Smirnov (KS) two sample test and the Gini coefficient, and eight features were selected for follow-up learning. Another method is to calculate the contribution level directly, e.g. Hassan et al. (2012) evaluated 20 features selected 9 of them.

Table 3. Scores for each feature contributes to the result

| feature           | score    |
| ----------------- | -------- |
| Pivot Energy      | 5.81E+05 |
| Variability Index | 1.87E+05 |
| Signif Avg        | 5.73E+02 |
| Signif Curve      | 1.57E+02 |
| PowerLaw Index    | 2.31E+01 |
| Spectral Index    | 2.07E+01 |
| Flux100 300       | 8.79E-06 |
| Flux300 1000      | 1.82E-06 |
| Flux1000 3000     | 1.38E-07 |
| Flux1000          | 8.53E-08 |
| Flux10000 100000  | 1.85E-08 |
| Flux Density      | 1.22E-08 |
| Energy Flux100    | 1.39E-09 |
| Flux3000 10000    | 4.07E-11 |

The Recursive Feature Elimination (RFE)[1] is a method that can evaluate the contribution of each feature to the classification label. In this work, it is carried out by using a package in *sklearn* ; the score of each feature is produced by the algorithm. The scores for all features are shown in Table 3. The contribution level of a feature to the classification result increases with the score.

---

[1] https://scikit-learn.org/stable/modules/generated/sklearn.feature selection.RFE.html

The above result differs from that of Kang et al. (2019). This indicates that different algorithms could produce different evaluation results. Thus, it seems that there is no common view on this issue. We apply principle component analysis (PCA) to resolve this issue and attempt to eliminate weak features.

3.3.2. *Principle Component Analysis*

Principal component analysis (PCA) is a statistical procedure that uses an orthogonal transformation to convert a set of observations of possibly correlated variables (original 14 observational features in this study) into a set of values of linearly uncorrelated variables (called the 'PCA result' in the following study) called principal components. Namely, if a specific coordinate has a large variance, more information is retained. Therefore, the related component is considered as a major one. In contrast, a smaller variance implies that the component might form through random fluctuation. Based on this idea, the contributions of small components are discarded to remove weakly relevant information. The PCA algorithm [1] we use can choose how many data characteristics to retain (from none, 0, to all, 1) to determine how many parameter dimensions to retain, or choose how many dimensions to retain directly. There four main steps of the PCA algorithm in this work are as follows:
(1) Generate a new dataset (denoted as the PCA result) by rotating and shifting the original coordinates. The generated components have the same dimensions as the original data, and this step does not cause any data loss.
(2) The explained variance for each dimension of the PCA result is calculated (automatically by the package). A larger variance indicates that more information is contained in this specific feature, see Table 5. For all the components, a larger explained variance value shows that it has a larger contribution. (Note: this step is an evaluation process carried out by the package, and no data loss occurs.)
(3) Discard the components which have an extremely small explained variance value. (Note: this step result in loss of data.)
(4) An inverse_PCA algorithm is applied to the reserved components, which returns the coordinates to their original states by rotating and shifting them. This produces a new set of 14-dimension data, denoted as 'new PCA result'. In this study, this 'new PCA result' will be used as a machine learning sample to minimise the loss of data .

Though this, we reduced the dimensions of the data based on the premise of retaining 99% of the data characteristics. As a result, the dimensions of the original data were reduced from 14 to 2 by the algorithm. This implies that there are only two major components in the original data, which correlates with the RFE result that the contribution scores of the first two features are significantly higher compared to those of the others.

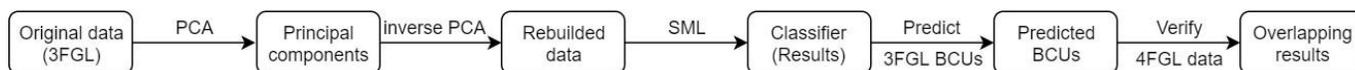

Figure 1. Data processing flow chart

We performed further verification: after extracting the entire Variability index feature (one of the two features with the highest contribution scores in the RFE result, see Table 3) from the original data, the PCA result showed that there was only one major component left. This indicates that our evaluation of contribution is consistent with the outcome of the PCA. We believe that our evaluation of contribution is accurate based on the consistency of the two methods. The result of dimension reduction (to 2-dimensional pricniple components) is shown in Fig 2.

After using PCA, BL Lacs and FSRQs have been roughly divided into two regions. However, a magnified view of the dense areas shows that there is overlap between numerous sources. This may indicate that a parameter space with only two dimensions cannot separate these two types of sources. Therefore, more parameter dimensions are required in constructing the PCA result. We reselected the data from the PCA result in Table 4 based on the explained variance of each component in Table 5.

---

[1] https://scikit-learn.org/stable/modules/generated/sklearn.decomposition.PCA.html?highlight=pcasklearn.decomposition.PCA

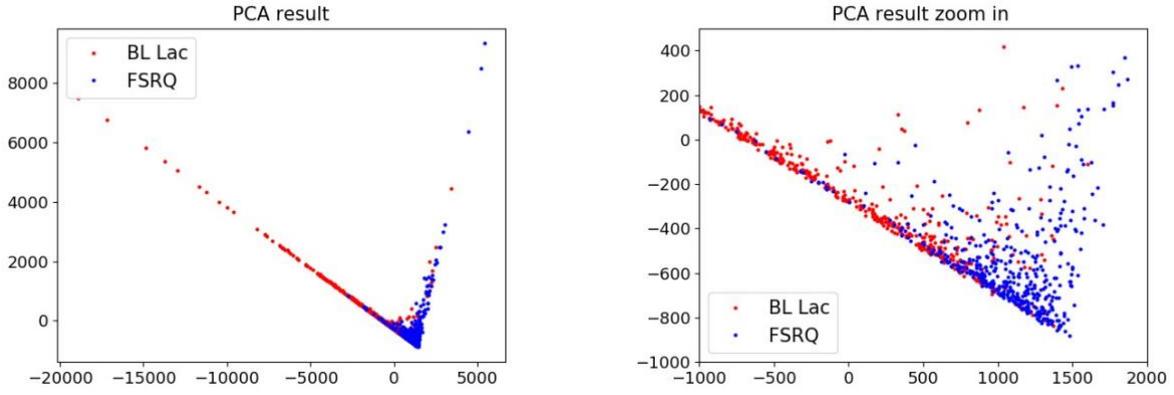

Figure 2. Result of PCA decomposition, the new two dimensional data is shown as left panel. The right panel is the magnified view of dense area of the result

Table 4. Eigenvalues of PCA result

| Source Name | 3FGL J0001.2-0748 | 3FGL J0001.4+2120 | 3FGL J0004.7-4740 | ... | 3FGL J2359.5-2052 |
|---|---|---|---|---|---|
| component 1 | 5.32E+02 | 1.44E+03 | 1.06E+03 | ... | -4.97E+02 |
| component 2 | -4.94E+02 | -7.82E+02 | -6.41E+02 | ... | -8.48E+01 |
| component 3 | -7.33E+00 | -9.79E+00 | -5.74E+00 | ... | -9.43E+00 |
| component 4 | -3.61E-01 | -2.81E+00 | -1.53E-01 | | 1.18E+00 |
| component 5 | -2.16E-01 | 1.32E-01 | 7.65E-02 | .. | -2.18E-01 |
| | | | | .. | |
| | | | | .. | |
| | | | | .. | |
| component 6 | -1.39E-02 | 2.87E-01 | -1.18E-02 | | 1.20E-02 |
| component 7 | -8.26E-10 | 2.05E-09 | -6.82E-09 | | 1.01E-08 |
| component 8 | -4.95E-10 | -2.03E-09 | -3.22E-10 | ... | 3.16E-10 |
| component 9 | -1.01E-10 | 1.11E-09 | -3.15E-10 | ... | -4.45E-11 |
| component 10 | 2.46E-11 | 5.03E-11 | 1.81E-11 | ... | -4.71E-11 |
| component 11 | -4.85E-11 | 2.65E-11 | -1.31E-11 | ... | 2.23E-11 |
| component 12 | -4.19E-11 | -9.54E-11 | -2.57E-11 | ... | 3.14E-11 |
| component 13 | 7.89E-12 | -3.25E-11 | -1.58E-12 | ... | 5.78E-12 |
| component 14 | 4.60E-13 | 9.62E-13 | 1.49E-14 | ... | -5.44E-13 |
| CLASS | bll | fsrq | fsrq | ... | bll |

Note—Each row in this table is denoted as 'PCA eigenvalue', this PCA result cause no any data reduction.

The first six explained variances of components are at least twelve orders of magnitude larger than the other components, which is consistent with the RFE evaluation results: there are six relatively evident components. Therefore, we believe that the first 6 components contribute most to the classification results, and thus we selected these components. It should be noted that the retention rate of data characteristics in the PCA result reaches 99% when only the first two components are selected. Hence, it can be assumed that there is almost no loss of data characteristics when the first six components are selected.

Finally, an inverse PCA algorithm was applied to the 6-dimensional PCA result to obtain a new, 14-dimensional dataset, denoted as 'new PCA features', and these data were used as the learning material in further machine learning. The accuracy and computing time of each model are shown in Table 2.

The results showed an improvement in the accuracy of all models except the SVM (linear) and SVM (RBF) and the KNN algorithm. In addition, computing time generally decreased, particularly for the MLP (L-BFGS) and SVM (linear). The most significant reduction in computing time was observed for LR.

Table 5. Explained variance of each PCA components

| PCA components | value |
| --- | --- |
| 1 | 4.84243386e+06 |
| 2 | 3.79797110e+06 |
| 3 | 3.86145730e+02 |
| 4 | 1.46478379e+00 |
| 5 | 7.18167051e-02 |
| 6 | 7.29951877e-04 |
| 7 | 3.89076727e-16 |
| 8 | 4.59159953e-18 |
| 9 | 4.26661328e-19 |
| 10 | 2.94054859e-20 |
| 11 | 2.54245910e-21 |
| 12 | 1.44362893e-21 |
| 13 | 3.23371252e-22 |
| 14 | 1.20176996e-24 |

3.3.3. *Analyses of cross validation against the testing set*

When the results of the testing set were compared with the results predicted by the model, all algorithms exhibited the same trend: they tended to identify BCUs as BL Lacs. For the labelled sources in 3FGL, the ratio of the number of BL Lacs and FSRQs is approximately 6:4(604 and 414), which indicates that BL Lacs are more likely to be observed than FSRQs. However, such imbalance in the learning materials might lead to the number of BL Lacs being significantly higher than the nunber of FSRQs in the learning results of all models.

It is worth noting that we apply a dichotomy for BCUs, that is, the result is either BL Lac or FSRQ, with the goal of definitively classifying ambiguous BCUs. In this study, all the default parameters of the package are used to build classifiers. The output probabilities of the RF, MLP, SVM and LR algorithms are float number ranging from 0 to 1, and the criterion is 0.5. However, in this package, the output probabilities of DT and KNN algorithms are either 0 or 1, so setting different criteria for these two classifiers is meaningless. This is unlike some other works, Salvetti et al. (2017) and Lefaucheur & Pita (2017) which divide the results into BL Lac, FSRQ, and even BCU. The proportion of the predicted BL Lacs and FSRQs differs from that in 3FGL: the result of every model contains more BL Lacs. This trend follows the result of Kang et al. (2019), which claims a proportion of 3:1. Therefore, classification as BL Lac may be more fault tolerant compared to FSRQ in the learning process, leading to this unbalanced result.

## 4. RESULTS AND DISCUSSION

We have studied the accuracy and computing time of different models with different kernels (when available). We attempted to use all original data to build the models; however, this was a time-consuming method of training the models. The RFE algorithm was used to evaluate the contribution level of each feature to reduce the number of features. Then, PCA was used to verify the results of the evaluation of the contribution level. The result shows that our evaluation was concordant with the PCA result. The PCA result was used to eliminate irrelevant or weakly relevant features to minimise the loss of data characteristics. Finally, the first 6 principle components from the result were selected. Then, the inverse PCA operation was applied to these principle components, and the resulting 14 dimensional 'new PCA result' were used as machine learning material.

### 4.1. *Classification result of BCUs*

Six machine learning algorithms with the highest accuracy were applied to establish the classification model, namely the MLP (L-BFGS), RF, the SVM (linear), DT, LR, and KNN. The predicted result for each model is shown in Table 6. Powerlaw index distribution histograms are plotted for the overall result (Fig.3) and for each model (Fig.4 and Fig.5), in which the labelled and predicted data are indicated in blue and orange, respectively. The distribution of the Powerlaw index of the predicted data shows good agreement with the labelled data. The results for the DT and KNN models produce an excess of BL Lacs on the right side of the distribution.

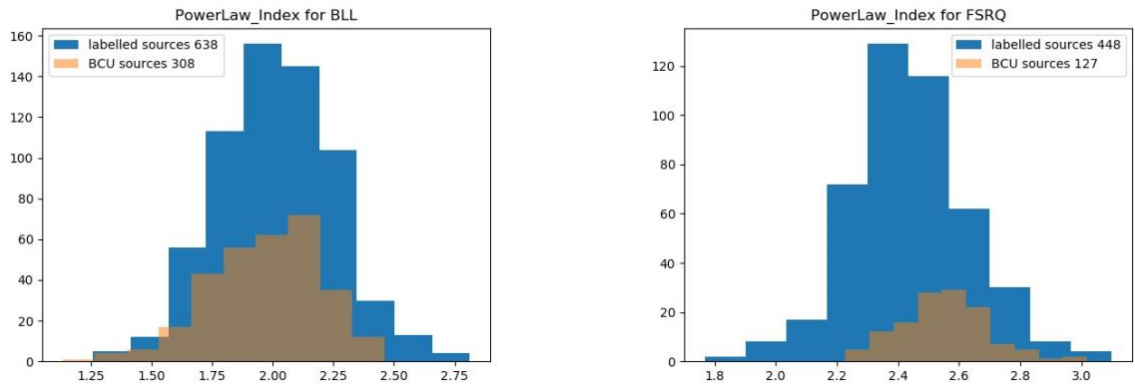

Figure 3. Left panel: distribution of power law index of labelled data and overlap of predicted data for BL Lac. Right panel: distribution of power law index of labelled data and overlap of predicted data for FSRQ

Table 6. Predicted results of each model for each BCU

| Source name | RF | SVM linear | MLP L-BFGS | LR | DT | KNN | overall[a] | 4FGL Name | 4FGL Class |
|---|---|---|---|---|---|---|---|---|---|
| 3FGL J0002.2-4152 | bll | bll | bll | bll | bll | bll | bll | 4FGL J0001.6-4156 | bcu |
| 3FGL J0003.2-5246 | bll | bll | bll | bll | bll | bll | bll | 4FGL J0003.1-5248 | bcu |
| 3FGL J0003.8-1151 | bll | bll | bll | bll | fsrq | bll | bcu | 4FGL J0003.9-1149 | bll |
| 3FGL J0009.6-3211 | bll | fsrq | bll | fsrq | bll | bll | bcu | 4FGL J0009.7-3217 | rdg |
| ... | ... | ... | ... | ... | ... | ... | ... | ... | ... |
| 3FGL J2353.7-3911 | bll | bll | bll | bll | fsrq | bll | bcu | 4FGL J2353.8-3911 | bcu |
| 3FGL J2358.3-2853 | bll | fsrq | bll | fsrq | fsrq | bll | bcu | 4FGL J2358.1-2853 | bcu |

$a$"overall" is the result of the overlap of six models

In terms of the number of predicted FSRQs, the trend of our classification results followed the result of (Yi et al. 2017), who classified BCUs according to morphology using a statistical method. In addition, our result is consistent with the result of (Salvetti et al. 2017), who used an artificial neural network (271 BL Lacs and 185 FSRQs). Our result is consistent with the results of (Chiaro et al. 2016) (342 BL Lacs and 154 FSRQs) and (Lefaucheur & Pita 2017) (417 BL Lacs and 149 FSRQs), who used a CNN to classify BCUs via a blazar flaring pattern. The numbers of predicted results of each model are shown in Table 7

Table 7. Amount of predicted results of each model

| | RF | SVM linear | MLP L-BFGS | LR | DT | KNN | overall[a] |
|---|---|---|---|---|---|---|---|
| bll | 377 | 438 | 386 | 351 | 392 | 414 | 308 |
| fsrq | 196 | 135 | 187 | 222 | 181 | 159 | 127 |

$a$"overall" is the result of the overlap of six models

## 4.2. *Performance of models*

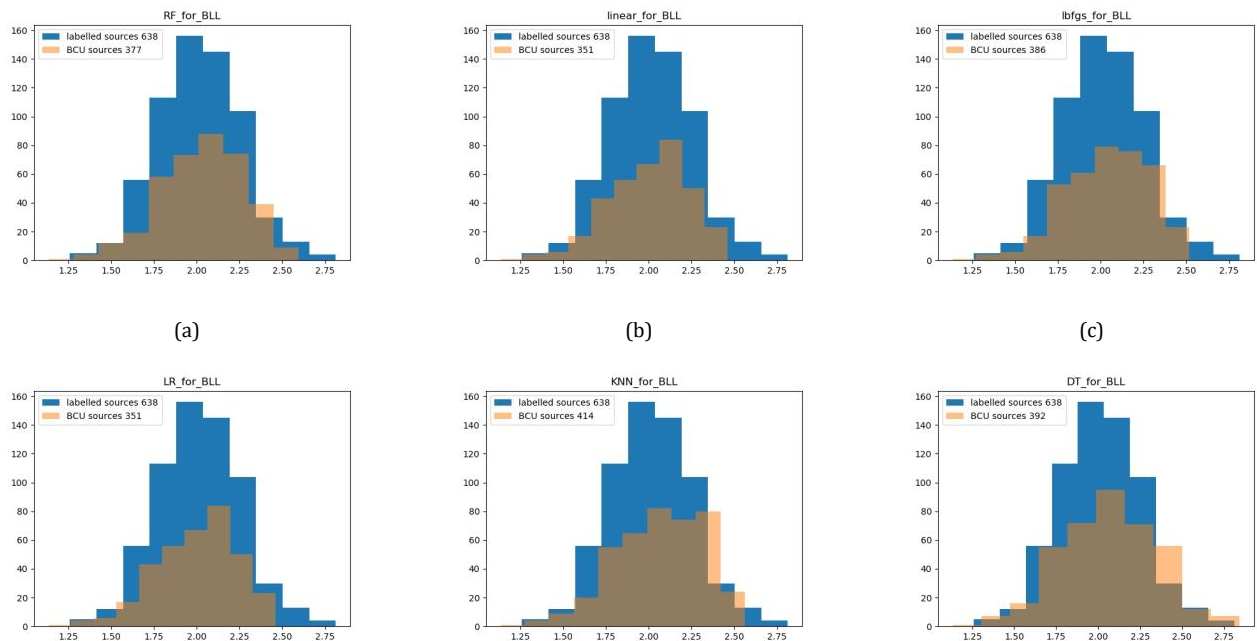

(a)  (b)  (c)

(d) (e) (f)

Figure 4. Distributions of power law index of labelled BL Lacs(blue) and predicted BL Lacs(orange) of each model. (a) RF, (b) SVM with linear kernel, (c) MLP with L-BFGS kernel, (d) LR, (e) KNN, (f) DT .

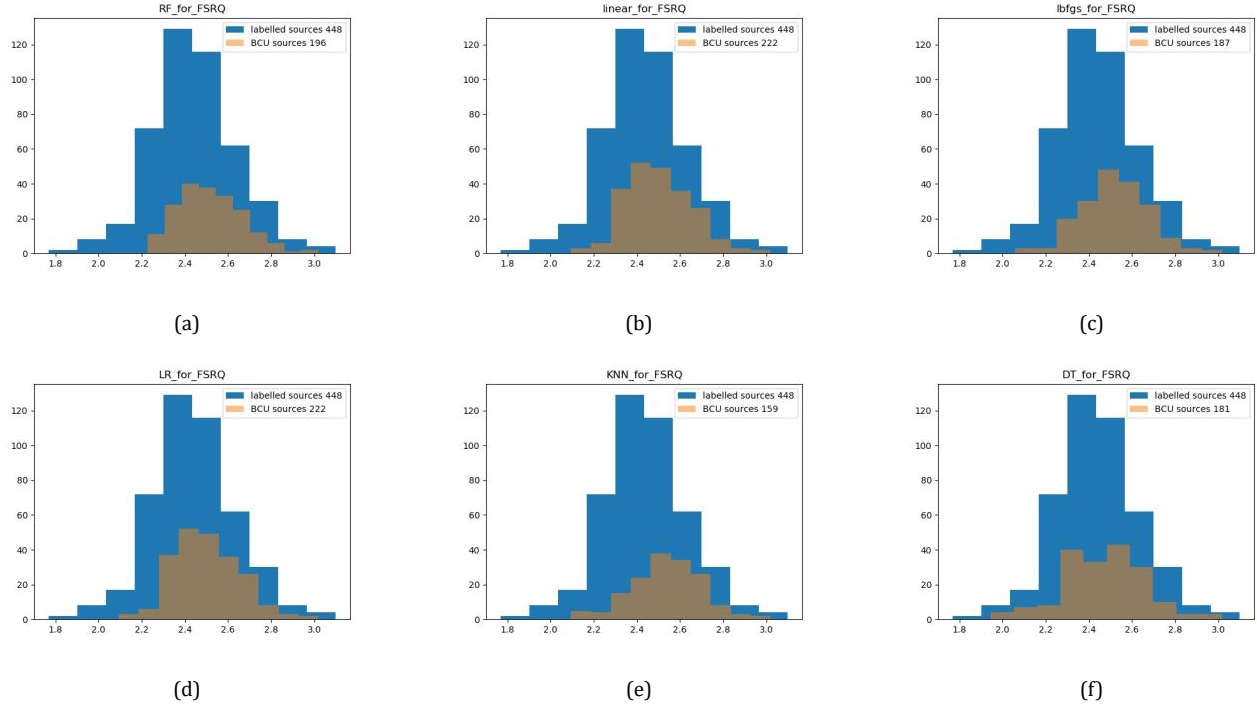

(a) (b) (c)

(d) (e) (f)

Figure 5. Distributions of power law index of labelled FSRQs(blue) and predicted FSRQs(orange) of each model. (a) RF, (b) SVM with linear kernel, (c) MLP with L-BFGS kernel, (d) LR, (e) KNN, (f) DT .

All six models perform well and have produced accurate results. Each specific classifier with its corresponding arguments is shown in Table 8. The accuracy, precision, true positive rate (TPR), and false positive rate (FPR) were calculated for each model and the results are shown in Table 9.

Table 8. Parameters of each model

| Model | function in Scikit-learn | arguments |
|---|---|---|
| RF | RandomForestClassifier() | n estimators=100, criterion='gini', max depth=None, min samples split=2, min samples leaf=1, max features='auto', max leaf nodes=None, bootstrap=True, max samples=None |
| SVM | SVC() | C=0.1, kernel='linear', degree=3, gamma='scale', coef0=0.0, shrinking=True, probability=False, tol=0.001, cache size=200, class weight=None, max iter=-1, decision function shape='ovr' |
| MLP | MLPClassifier() | solver='lbfgs', activation='relu', alpha=0.011, batch size=200, learning rate=0.001, max iter=200, tol=1e-4 |
| LR | LogisticRegression() | penalty='l2', dual=False, tol=0.0001, C=1.0, fit intercept=True, intercept scaling=1, solver='lbfgs', max iter=1000, multi class='auto' |
| DT | DecisionTreeClassifier() | criterion='gini', splitter='best', max depth=None, min samples split=2, min samples leaf=1, max features=None, random state=None, max leaf nodes=None, presort='deprecated' |
| KNN | KNeighborsClassifier() | n neighbors=5, weights='uniform', algorithm='auto', leaf size=30, p=2, metric='minkowski', metric params=None |

Moreover, the computing times of almost all algorithms were reduced, particularly the LR algorithm, which requires less than 6.5 s, see Table 2.

Table 9. Accuracy, Precision, TPR and FPR of each model

| | 3FGL | | | | 4FGL | | | |
|---|---|---|---|---|---|---|---|---|
| Model | Accuracy | Precision | TPR | FPR | Accuracy | Precision | TPR | FPR |
| RF | 91.28% | 84.43% | 91.26% | 17.92% | 89.74% | 92.94% | 95.18% | 32.00% |
| SVM | 89.45% | 93.20% | 85.71% | 6.60% | 87.18% | 88.82% | 96.18% | 24.00% |
| MLP | 92.20% | 94.39% | 90.18% | 5.66% | 89.23% | 92.35% | 95.15% | 32.00% |
| LR | 90.83% | 93.33% | 87.50% | 6.60% | 87.18% | 88.82% | 96.18% | 24.00% |
| DT | 86.70% | 84.60% | 91.07% | 21.70% | 85.13% | 88.24% | 94.34% | 36.00% |
| KNN | 87.16% | 83.90% | 88.39% | 17.92% | 91.28% | 95.29% | 94.74% | 36.00% |

The precisions of RF, SVM, MLP, and LR are higher than 90%. However, the precisions of DT and KNN are relatively low, which may be caused by their higher FPRs (21.70% and 17.92%, respectively), indicating that these models may have a inferior ability to classify BCUs. The classification ability of each model is significant and merits further evaluation and discussion. For this purpose, an ROC curve[3] (Fawcett 2006) (Fig.6) is plotted to evaluate the classification responsiveness of a classifier for each model. A larger area under the curve indicates a superior classification ability of the model is better. Additionally, a smoother curve lessens the opportunity to overfit the data. The areas under the ROC curves of the six models are 0.95 for RF, 0.95 for SVM, 0.96 for MLP, 0.95 for LR, 0.87 for DT, and 0.87 for KNN. The area under each curve and their shapes indicate that the RF, SVM, MLP, and LR models possess better performance compared to the DT and KNN models. This is consistent with the excess BL Lacs in the histogram of the spectral index distribution of these two models.

[3] https://scikit-learn.org/stable/modules/generated/sklearn.metrics.roc curve.html?highlight=rocsklearn.metrics.roc curve

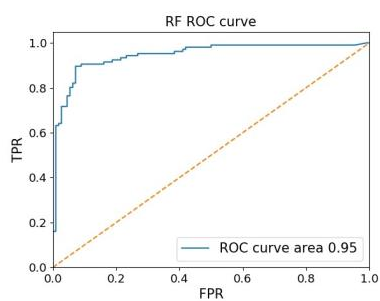
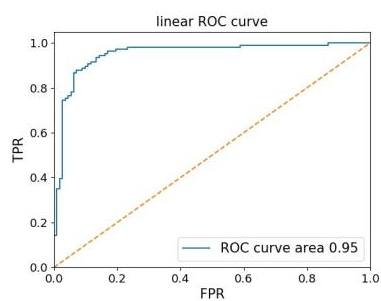
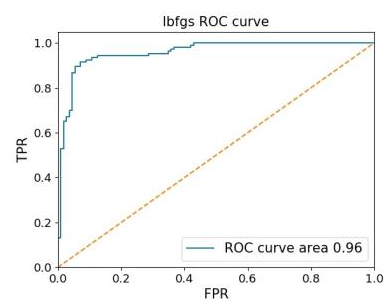

(a)        (b)        (c)

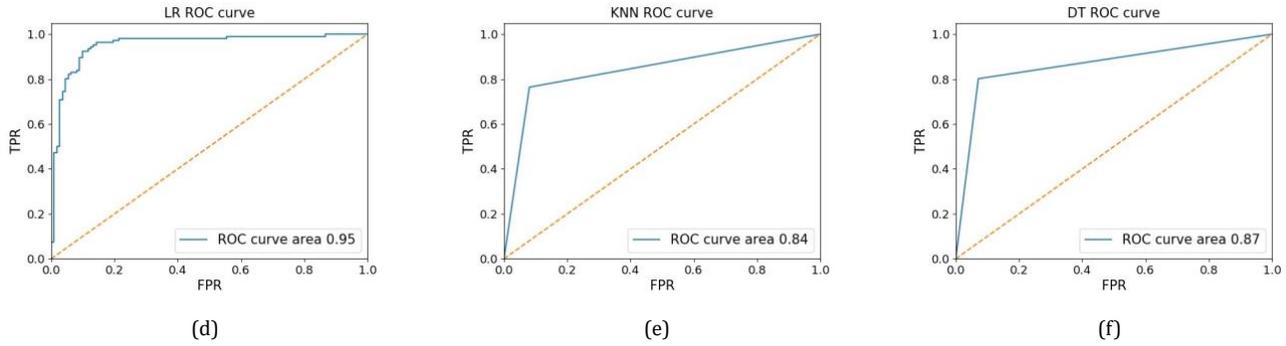

Figure 6. ROC curve for each model. (a) RF, (b) SVM with linear kernel, (c) MLP with lbfgs kernel, (d) LR, (e) KNN, (f) DT .

### 4.3. *Verification of the classification result of BCUs*

The result of each model shows a good agreement with 4FGL, indicating that our hybrid method can indeed produce an accurate classification result. Some of the BCUs in 3FGL have been identified in 4FGL, among the 573 BCUs in 3FGL, 171 were identified as BL Lacs, 24 as FSRQs, and 326 remain as BCUs; the others were unlabelled or have been re-categorised as other types of sources. Traditionally, the accuracy of each classifier is evaluated from the test set which is still a part of the known learning material. This was carried out as a significant evaluation result to evaluate the performance of classifiers in previous works. However, as a newgeneration catalog with true values, the 4FGL data could further help evaluation of the performance of each model. To verify whether our predicted results are correct, they were compared with the associated ones in 4FGL(Table 6) and the overlapping results are listed in Table 10. Furthermore, the accuracy, precision, TPR and FPR of each classifier are calculated from the result overlapping with the 4FGL data and are shown in Table 9. From the overlapping results, KNN, RF, and MLP show a better performance, correctly classifying over 90% of identified BCUs in 4FGL. Evaluation results calculated via 4FGL are consistent with those calculated via the test set, indicating that this hybrid method is more reliable and effective.

Table 10. Comparing results of six models and 4FGL

|         | RF     | SVM    | MLP    | LR     | DT     | KNN    | LP[a]  | C[b]   |
|---------|--------|--------|--------|--------|--------|--------|--------|--------|
| BLL     | 158    | 151    | 157    | 151    | 150    | 162    | 142    | 151    |
| BLL/171 | 92.40% | 88.30% | 91.81% | 88.30% | 87.72% | 94.74% | 83.04% | 88.30% |
| FSRQ    | 17     | 19     | 17     | 19     | 16     | 16     | 12     | 13     |

*Table 10 continued on next page*

Table 10 *(continued)*

|         | RF     | SVM    | MLP    | LR     | DT     | KNN    | LP[a]  | C[b]   |
|---------|--------|--------|--------|--------|--------|--------|--------|--------|
| FSRQ/24 | 58.33% | 79.17% | 70.83% | 79.19% | 66.67% | 66.67% | 50%    | 54.17% |

[a]Result predicted by Lefaucheur & Pita (2017) [b]Result predicted by Chiaro et al. (2016)

For comparison, the results of Lefaucheur & Pita (2017) , Chiaro et al. (2016) and Kang et al. (2019) are compared with the associated results in 4FGL. It is necessary to note that the result of Kang et al. (2019) is obtained on a clean sample. The sources that were not present in the clean sample were eliminated from our result to generate a new 'clean result'. The comparison of the results is shown in Table 11. The percentages in Table 10 and Table 11 indicate the ratios between correctly identified BCUs and all the identified BCUs (171 BL Lacs and 24 FSRQs in Table 10, 141 BL Lacs and 20 FSRQs in Table 11).

Table 11. Comparing results of six models and 4FGL on clean sample

|  | RF | SVM | MLP | LR | DT | KNN | Kang[a] |
|---|---|---|---|---|---|---|---|
| BLL | 132 | 126 | 131 | 127 | 126 | 135 | 84 |
| BLL/141 | 93.62% | 89.36% | 92.91% | 90.07% | 89.36% | 95.74% | 59.57% |
| FSRQ | 11 | 13 | 11 | 13 | 11 | 11 | 7 |
| FSRQ/20 | 55% | 65% | 55% | 65% | 55% | 55% | 35% |

[a]The result of Kang et al. (2019) is on https://github.com/ksj7924/Kang2019ApJRcode

### 4.4. *Discussion*

The good agreement of the above comparison indicates that removing irrelevant or weakly relevant features through PCA can produce a precise classification result. Blazars can be divided into BL Lacs or FSRQs based on the differences in their optical spectra, and this is one of the main methods of classifying blazars. Even if other observational features cannot classify the source directly, they may contain potentially useful classification information. In the detected spectrum range of Fermi LAT in the gamma-ray band, BL Lacs are generally believed to be the result of synchron self-Compton (SSC) emission, whereas FSRQs are cause by SSC emission and an External Compton scattering. These differing mechanisms complicate the observational data from FSRQs. Furthermore, the spectrum of FSRQs in the Fermi range can include contributions from other components. This suggests that, for FSRQs, the detected data from Fermi might represent other more complicated physical processes.

Discarding features to improve accuracy may cause loss of data characteristics, which may obfuscate inner physics features. Therefore, even though a few features have a relatively lower contribution to the classification result, we do not discard them directly. Instead, we apply the PCA algorithm to tune the data by reducing its dimension while maximally for preserving data characteristics, thereby reducing computing time and improving accuracy.

The most important part of this work is that we utilise an updated dataset(4FGL) as a true value to verify the predicted result(BCUs in 3FGL), rather than verifying the result from a test set of 3FGL. A higher ratio of overlapping between predicted BCUs and 4FGL data indicates that this hybrid method is more effective and generalised. Both the result obtained by combining PCA and SML, and the results of some previous works which were obtained directly from SML, have a high accuracy near 90%. However, the hybrid method in this study shows better agreement with 4FGL. This might indicates that the hybrid method can produce a more reliable and more generalised classification result. Selection effects of the instruments, e.g. detection threshold, can also affect such work, especially in terms of the unbalanced number of BL Lacs and FSRQs in the 3FGL.

The results we have obtained from this work, especially the overlap result (*overall* column in Table 6), can serve as reference for forthcoming surveys or other ground based spectral observation. The enlarged database containing the new classified BCUs could be used in future research, especially statistical research. If the accuracy of evaluating Fermi data using these kinds of SML methods could be further improved in the future, it could be used as a brand-new defining method based on the gamma ray waveband.

Salvetti et al. (2017) suggested that if a new subclass of blazars was created, the number of blazars that remain unclassified in their work could be reduced from 52% to 10%. This suggests that the current classification standard may cover up certain inner physics features, leading to some sources retaining the label of BCU.

Additionally, in the field of machine learning, the data we used in this study might be considered a small sample, even though all the labelled data in 3FGL is used in this work. Nonetheless, it was still sufficient for building the classifiers through SML. Obtaining the accuracy and precision for each model might appear to be an unrealistically high precision. However, as a predicted result which are used as a referee for further observations, the classification result of these unknown data obtained a high accuracy while verified with 4FGL that includes the true value as those unknown data. This indicates the classification result of this hybrid method is reasonable. The good agreement between the classification result and the test set may contain bias from the entire sample, so comparison with another database is necessary.

## 5. CONCLUSION

Astronomy is a science based on observation. Our aim is to not only classify the BCUs in 3FGL accurately, but also to present a new approach for processing data. In the field of machine learning, reducing the amount of irrelevant information in a training set is a common method of improving performance. However, there are different types of methods of evaluating contribution of features. Therefore, we do not reduce the numbers of features to decrease potential bias; instead, we apply PCA to tune it. We obtain a series of BL Lacs and FSRQs after cross validating the results obtained from different models. This can provide a reference for selecting sources for follow-up observation to a certain extent.

Compared to previous works, we did not evaluate the reliability and accuracy of models in various aspects but crossvalidated the data in 4FGL; this can be regarded as a true classification result. This indicates that the combination of machine learning and PCA can effectively classify blazars into 2 subclasses. The observed agreement with 4FGL indicates that our models have reduced bias and overfitting, and a high universal applicability.

All data and processing scripts can be download from https://github.com/astronomical-data-processing/ bcus-classification.

We thank all the anonymous referee for every valuable and helpful comments and suggestions. This work is based on the published data from Fermi Satellite data base. This work is supported by the National Key Research and Development Program of China (2018YFA0404603), the Joint Research Fund in Astronomy (U1831204, U1931141) under cooperative agreement between the National Natural Science Foundation of China (NSFC) and the Chinese Academy of Sciences (CAS), Funds for International Cooperation and Exchange of the National Natural Science Foundation of China (11961141001), National Science Foundation for Young Scholars (11903009). Yunnan Key Research and Development Program(2018IA054). The major scientific research project of Guangdong regular institutions of higher learning (2017KZDXM062). This work is also supported by Astronomical Big Data Joint Research Center, co-founded by National Astronomical Observatories, Chinese Academy of Sciences and Alibaba Cloud, the Innovation Research for the postgraduates of GuangZhou University (2019GDJC-M46).

## REFERENCES

Abdollahi, Acero, A. 2019, arXiv preprint arXiv:1902.10045

Acero, F., Ackermann, M., Ajello, M., et al. 2015, The Astrophysical Journal Supplement Series, 218, 23

Arthur, D., & Vassilvitskii, S. 2006, k-means++: The advantages of careful seeding, Tech. rep., Stanford

Bai, Y., Liu, J.-F., & Wang, S. 2018, Research in Astronomy and Astrophysics, 18, 118

Ball, N. M., & Brunner, R. J. 2010, International Journal of Modern Physics D, 19, 1049

Breiman, L. 2001, Machine learning, 45, 5

Chiaro, G., Salvetti, D., La Mura, G., et al. 2016, Monthly Notices of the Royal Astronomical Society, 462, 3180

Einecke, S. 2016, Galaxies, 4, 14

Fawcett, T. 2006, Pattern Recognition Letters, 27, 861

Hassan, T., Mirabal, N., Contreras, J., & Oya, I. 2012, Monthly Notices of the Royal Astronomical Society, 428, 220

Kang, S.-J., Fan, J.-H., Mao, W., et al. 2019 (American Astronomical Society), 189. https://doi.org/10.3847%2F1538-4357%2Fab0383

Lefaucheur, J., & Pita, S. 2017in , AIP Publishing, 050024

Mirabal, N., Frias-Martinez, V., Hassan, T., & Frias-Martinez, E. 2012, Monthly Notices of the Royal Astronomical Society: Letters, 424, L64

Parkinson, P. S., Xu, H., Yu, P., et al. 2016, The Astrophysical Journal, 820, 8

Pedregosa, F., Varoquaux, G., Gramfort, A., et al. 2011, Journal of Machine Learning Research, 12, 2825 Rumelhart, D. E., Hinton, G. E., & Williams, R. J. 1985, Learning internal representations by error propagation, Tech. rep., California Univ San Diego La Jolla Inst for Cognitive Science

Salvetti, D., Chiaro, G., La Mura, G., & Thompson, D. J. 2017, Monthly Notices of the Royal Astronomical Society, 470, 1291

Utgoff, P. E. 1989, Machine learning, 4, 161

Vapnik, V. 2013, The nature of statistical learning theory (Springer science & business media)

Yi, T.-F., Zhang, J., Lu, R.-J., Huang, R., & Liang, E.-W. 2017, The Astrophysical Journal, 838, 34